\documentclass[useAMS,usenatbib]{mn2e}

\usepackage{epsfig}

\def\llm{{\sc LLmodels}}
\def\atl{{\sc ATLAS9}}
\def\width9{{\sc WIDTH9}}
\def\detail{{\sc DETAIL}}
\def\surface{{\sc SURFACE}}
\def\binmag{{\sc BinMag}}
\def\templogg{{\sc TempLogG$^{\rm TNG}$}}
\def\vald{{\sc VALD}}
\def\logg{\log g}
\def\teff{T_{\rm eff}}

\def\kms{km/s}
\def\halpha{\mathrm{H}\alpha}
\def\hbeta{\mathrm{H}\beta}

\def\ddafit{{\sc DDAFit}}
\def\synth3{{\sc SYNTH3}}
\def\ei{E_{\rm i}}

\def\loggf{\log(gf)}
\def\vsini{\upsilon\sin i}

\def\Rsun{R_{\odot}}
\def\gammas{\gamma_\mathrm{Stark}}
\def\etal{et al.}
\def\mh{\mathrm{[M/H]}}

\def\C   {C\,{\sc  i}}  
\def\O   {O\,{\sc  i}} 
\def\Na  {Na\,{\sc i}} 
\def\Mgi {Mg\,{\sc i}} 
\def\Mgii{Mg\,{\sc ii}}
\def\Al  {Al\,{\sc ii}}
\def\Sii {Si\,{\sc i}} 
\def\Siii{Si\,{\sc ii}}
\def\Ca  {Ca\,{\sc ii}}
\def\Ti  {Ti\,{\sc ii}}
\def\V   {V\, {\sc ii}} 
\def\Cri {Cr\,{\sc i}} 
\def\Crii{Cr\,{\sc ii}}
\def\Mni {Mn\,{\sc i}} 
\def\Mnii{Mn\,{\sc ii}}
\def\Fei {Fe\,{\sc i}} 
\def\Feii{Fe\,{\sc ii}}
\def\Co  {Co\,{\sc ii}}
\def\Ni  {Ni\,{\sc i}} 
\def\Ba  {Ba\,{\sc ii}}
\def\Ce  {Ce\,{\sc ii}}
\def\Pr  {Pr\,{\sc iii}}
\def\Nd  {Nd\,{\sc iii}}
\def\Sm  {Sm\,{\sc ii}}
\def\Eu  {Eu\,{\sc ii}}
\def\Gd  {Gd\,{\sc ii}}

\def\hd{HD~103498}

\title[Abundance and stratification analysis of the CP star HD~103498]
{Abundance and stratification analysis of the CP star HD~103498}
\author[C. P. Pandey, D. V. Shulyak, T. Ryabchikova and O. Kochukhov]
{Chhavi P. Pandey$^{(1,2)}$\thanks{chhavi.pandey@gmail.com}, Denis V. Shulyak$^{(3)}$, Tanya Ryabchikova$^{(4)}$ and  \newauthor   Oleg Kochukhov$^{(5)}$\\
$^{(1)}$Aryabhatta Research Institute of Observational Sciences, Nainital, 263129, India \\
$^{(2)}$Department of Physics, Kumaun University Nainital - 263002, India. \\
$^{(3)}$Institute of Astrophysics, Georg-August-University, Friedrich-Hund-Platz 1, D-37077 G\"ottingen, Germany \\
$^{(4)}$Institute of Astronomy, Russian Academy of Sciences, Pyatnitskaya 48, 119017 Moscow, Russia \\
$^{(5)}$Department of Physics and Astronomy, Uppsala University, Box 516, 751 20, Uppsala, Sweden}

\begin{document}

\date{}

\pagerange{\pageref{firstpage}--\pageref{lastpage}} \pubyear{2011}

\maketitle

\label{firstpage}

\begin{abstract}
Slow rotation and absence of strong mixing processes in atmospheres of chemically peculiar stars develop ideal conditions for the appearance of abundance anomalies through the mechanism of microscopic particle diffusion. This makes these objects look spectroscopically and photometrically different from their ``normal'' analogs. As a result, it is often difficult to accurately determine atmospheric parameters of these stars and special methods are needed for the consistent analysis of their atmospheres. \\
The  main  aim of the present paper is to analyse atmospheric abundance and stratification of chemical elements in the atmosphere of the chemically peculiar star \hd.\\
We find that two model atmospheres computed with individual and stratified abundances provide reasonable fit to observed spectroscopic and photometric indicators: $\teff=9300$~K, $\logg=3.5$ and $\teff=9500$~K, $\logg=3.6$. It is shown that Mg has a large abundance gradient in the star's atmosphere with accumulation of Mg ions in the uppermost atmospheric layers, whereas Si demonstrates opposite behaviour with accumulation in deep layers. In addition, a  detailed non-LTE analysis showed that none of Mg transitions under consideration is a subject of noticeable non-LTE effects. Comparing observed photometry transformed to physical units we estimated the radius of \hd\ to be between $R=(4.56\pm0.77)\Rsun$ for $\teff=9300$~K, $\logg=3.5$ and $R=(4.39\pm0.75)\Rsun$ for $\teff=9500$~K, $\logg=3.6$ models respectively. We note that the lack of suitable observations in absolute units prevents us to uniquely determine the $\teff$ of the star at the current stage of analysis.

\end{abstract}


\section{Introduction}\label{intro}
As name indicates, Chemically Peculiar (CP) stars are unusual in chemical composition in comparison to that 
of normal stars with similar fundamental parameters.  This family of stars resides on the upper main sequence 
and they contribute about $15\% - 20\%$ of early B to late F-type stars. The speciality of this group is that 
they hold the strong abundance anomalies in their atmospheres such as non-uniform horizontal and vertical 
distributions of chemical elements, and/or surface magnetic fields of different intensities. 
The type of peculiarity varies from star to star and it depends upon many factors like:
stellar effective temperature, rotational velocity, the presence of magnetic field, 
membership in binary stellar systems etc. Microscopic chemical diffusion, arising from a competition 
between radiative levitation and gravitational settling is believed to be the main process behind this 
phenomena, as firstly discussed by \citet{michaud}.

\hd\ is a characteristic example of an evolved star belonging to the co-called CP group of spectral type A (Ap) with strong overabundance of Si, Fe, and Cr in its atmosphere as follows from the recent analysis undertaken by \citet{Joshi}. The authors aimed on the pulsation analysis of the star's atmosphere and performed only classical homogeneous abundance analysis based on scaled-solar metallicity model atmosphere. Their analysis did not provide clear evidence whether \hd\ belongs to roAp (rapidly-oscillating Ap) stellar group. They also noted that the chemical properties and a high effective temperature of the star are distinguishable among known roAp stars. According to the recent spectropolarimetric observations of \cite{am} this star has a weak longitudinal field $<B_{z}>$ that varied between $\pm$200 G with a period of 15.83 days.

In this work we extend the study of \citet{Joshi} and carry out complex and self-consistent analysis
of the \hd\ atmosphere. The main goal is to involve all the available observables to derive consistently
fundamental atmospheric parameters and abundance pattern of the star. We also look for stratification of 
chemical elements and attempt to construct a self-consistent atmospheric model  which fits metallic line spectra, hydrogen line profiles, and energy distribution of the star. This approach where an iterative process is used for the derivation of atmospheric parameters was already successfully applied in a number of recent investigations
of normal \citep{2009A&A...506..203R} and peculiar \citep[e.g.][]{2009A&A...499..879S,2010A&A...520A..88S,2009A&A...499..851K} stars.
The case of \hd\ is of certain interest also because of its relatively high (as for roAp stars) 
temperature ($\teff\approx9500$~K) at which such important spectroscopic indicators as hydrogen lines become sensitive both to the atmospheric temperature and pressure stratification making it difficult to separate effects of $\teff$ and $\logg$. Thus other (photometric and/or spectrophotometric) observables must be simultaneously involved in the analysis to obtain precise results.


 \section{Observations}
We used the average spectrum of \hd\ which were obtained on February 2, 2009 (HJD 2454865.624--2454865.764) 
using the 2.56-m Nordic Optical Telescope (NOT) equipped with the Fibre-fed Echelle Spectrograph (FIES).
The instrument was configured to use the high-resolution mode at the resolving power of $R=47\,000$ in the spectral region  
$\lambda\lambda~3900-7270$. The FIES spectra  were acquired at the magnetic phase range $0.37\pm0.2$. 
The abundance analysis of the program star is based on the average spectrum of time-series spectroscopy.  
The finer details about the observations and data reduction procedure are given in \citet{Joshi}. 

We also made use of available photometric observations in Geneva \citep{1976A&AS...26..275R}, 
Johnson \citep{1973PASP...85...85H}, Str\"omgren \citep{1998A&AS..129..431H}, 2MASS \citep{2006AJ....131.1163S},
and $\Delta~a$ \citep{2005A&A...441..631P} systems.


 \section{Analysis tools and methodology}

\subsection{Model atmospheres}
To perform the model atmosphere calculations we used the recent version of the \llm\, \citep{llm} 
stellar model atmosphere code. \llm\, is 1-D, hydrostatic, plain-parallel LTE code which accounts for the effects of 
individual and stratified abundances. This program treats the bound-bound opacity by direct, 
line-by-line spectrum synthesis. The stratification of chemical elements is an input parameter 
for the code and thus is not changing during model atmosphere calculation process. 
This allows to explore the changes in model structure due to stratification that was extracted directly from observations  
without modeling the processes that could be responsible for the observed inhomogeneities. 

For every model atmosphere calculation the stellar atmosphere is discretized into $120$ layers between 
$\log \tau_{\rm 5000} = -8$ and $2$ with the higher point density in the region of steep abundance gradients 
to ensure accurate integration of radiation field properties and solution of other equations. 
The frequency-dependent quantities are calculated with variable wavelength step with a total of
$\approx520\,000$ wavelength  points spread over the spectral region where the star
radiates most of its energy. 
For the  computation of line absorption coefficient \vald\ database \citep{piskunov, kupka} 
was used as a source of atomic lines data.

\subsection{Classical abundance analysis} 

The  line identification in the observed spectrum of the star is based on the theoretical spectrum calculated 
for the  entire observed spectral region using lines extracted from \vald. The spectrum synthesis code \synth3\ \citet{ddafit07} was used 
in all synthetic spectrum calculations.
We compared synthetic and observed spectra to choose the least blended lines for the accurate abundance analysis 
using IDL package \binmag\footnote{http://www.astro.uu.se/$\sim$oleg/}.

The classical abundance analysis in the approximation of chemically-homogeneous atmosphere is based on the analysis
of equivalent  widths of the lines, which is performed using the updated version \citep{vadim} of \width9\ code \citep{a9-2}. 
We adopted $\vsini = 12$~\kms\ (obtained by the best fit to the observed unblended line profiles) and microturbulent velocity $\xi= 1.0 \pm 0.2$~\kms\ (derived from numerous lines of \Cri , \Crii , \Fei\   and \Feii) from \cite{Joshi}. As atmospheres of CP stars of type A are believed to be quiet in order to develop chemical peculiarities, a non-zero microturbulence found in \citep{Joshi} probably results from the Zeeman broadening of spectral lines caused by a weak surface magnetic field. Since both in Joshi et al and in the present study we use non-magnetic spectra synthesis, $\xi= 1.0$~\kms\ was still needed to mimic the effect of the star's magnetic field and its value were kept constant during iterative abundance analysis described below.

\subsection{Stratified abundance analysis}
\label{sec:strat}

To study the stratification of chemical elements in the atmosphere of the star we used the \ddafit\ script which is 
an automatic procedure to find chemical abundance gradients from the observed spectra \citep{ddafit07}. 
\ddafit\ provides an optimization and visualization interface to the spectrum synthesis calculations. 
In this routine, the vertical abundance distribution of an element is described with the four parameters: 
chemical abundance in the upper atmosphere, abundance in deep layers, the vertical position of abundance step 
and the width of the transition region where chemical abundance changes between the two values. 
All four parameters can be modified simultaneously with the least-squares fitting procedure and based on 
observations of unlimited number of spectral regions. 
This procedure is successfully applied in a number of studies \citep[e.g.][]{tr3,tr2,tr1,2009A&A...499..851K,2009A&A...499..879S,2010A&A...520A..88S}.

\subsection{Non-LTE analysis codes}

To analyse the non-LTE effects on the \Mgi/\Mgii\ ions we used the \detail\ and \surface\ codes originally developed by Butler and Giddings \citep{butler,giddings}. Our calculations take into account the recent improvements in the atomic data for Mg. The extensive description of its model atom and non-LTE line formation are presented in \citet{przybilla} and we refer the interested reader to the aforementioned paper.

\subsection{Self-consistent atmospheric modeling}
\label{sec:iter}

Using the analysis techniques outlined above, it is possible to reconstruct the element distribution profile of any chemical species for which the accurate atomic data exist. However it is to note that the empirical analysis  of chemical elements stratification depends upon model atmosphere used, and the temperature-pressure structure of model atmosphere itself relies upon stratification. Consequently, the model atmosphere construction and stratified abundance analysis both are strongly coupled together. It is therefore necessary to use the iterative procedure which is described in different steps as follows:
\begin{enumerate}
\item selection of atmospheric parameters (first approximation, i.e. initially estimated  parameters must be close to those
of the star to be analysed; they can be either taken from the literature or estimated using available photometric calibrations);
\item 
analysis of individual spectral lines to determine the stratification of chemical elements;
\item
with the help of stratification found in previous step, new model must be calculated to provide improved 
temperature-pressure structure of the stellar atmosphere;
\item
comparison of the  modeled energy distribution (or/and photometric colors) and hydrogen line profiles 
with the observed ones. The  model input parameters ($\teff$, $\logg$) are then tuned to provide an agreement between
theory and observations.
\item
finally, the overall process from step 2 onwards must be repeated until stratification profiles of chemical 
elements and model parameters are converged.
\end{enumerate}

\begin{figure}
\begin{center}
\epsfig{file=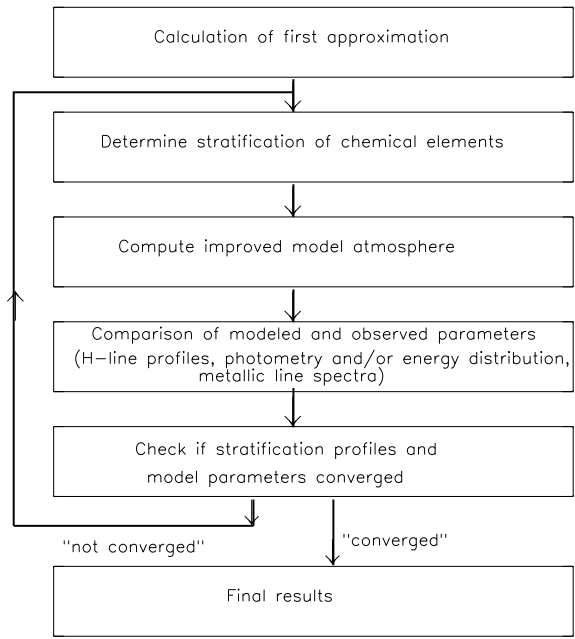,width=.6\textwidth}
\caption{Steps used for the self-consistent models with empirical abundance/stratification analysis.}
\label{flow}
\end{center}
\end{figure}

The mentioned above  steps are schematically presented in Fig.~\ref{flow}. 
This procedure guarantees the  consistency  between the atmospheric model structure and abundances used for the calculation of synthetic line  profiles.

The final self-consistent, chemically-stratified model atmosphere is expected to reproduce simultaneously the observed 
photometry, energy distribution, hydrogen line profiles and metallic line spectra. Some recent examples of 
applying this procedure can be found in \citet{2009A&A...499..851K} and \citet{2009A&A...499..879S,2010A&A...520A..88S}.
 
\begin{table}
\caption{A list of spectral lines used for the stratification analysis.}
\label{Tstrat-list}
\begin{footnotesize}
\begin{center}
\begin{tabular}{lcrrrl}
\noalign{\smallskip}
\hline
Ion & Wavelength &$\ei$,(eV)  & $\loggf$ & $\log\gammas$ & Ref.\\
\hline
\Mgii& 4384.637 &  9.996 & $-$0.790 &  $-$4.07 & KP \\
\Mgii& 4390.514 &  9.999 & $-$1.490 &  $-$4.07 & KP \\
\Mgii& 4390.572 &  9.999 & $-$0.530 &  $-$4.07 & KP \\
\Mgii& 4433.988 &  9.999 & $-$0.910 &  $-$4.40 & KP \\
\Mgii& 4481.126 &  8.864 &    0.740 &  $appx$  & KP \\
\Mgii& 4481.150 &  8.864 & $-$0.560 &  $appx$  & KP \\
\Mgii& 4481.325 &  8.864 &    0.590 &  $appx$  & KP \\
\Mgi & 4702.991 &  4.346 & $-$0.666 &  $-$3.98 & LZ \\
\Mgii& 4739.593 & 11.569 & $-$1.960 &  $appx$  & KP \\
\Mgii& 4739.593 & 11.569 & $-$0.660 &  $appx$  & KP \\
\Mgii& 4739.709 & 11.569 & $-$0.820 &  $appx$  & KP \\
\Mgi & 5172.684 &  2.712 & $-$0.402 &  $appx$  & AZ \\
\Mgi & 5183.604 &  2.717 & $-$0.180 &  $-$7.99 & AZ \\
\Mgi & 5528.405 &  4.346 & $-$0.620 &  $-$4.46 & LZ \\
\Mgi & 5711.088 &  4.346 & $-$1.833 &  $appx$  & LZ\\\hline
\Siii& 4130.872 &  9.839 & $-$0.824 &  $-$4.87 & BBCB \\
\Siii& 4190.707 & 13.492 & $-$0.17  &  $-$5.26 & MERL \\
\Siii& 5055.984 & 10.074 &    0.593 &  $-$4.78 & SG   \\ 
\Siii& 5056.317 & 10.074 & $-$0.359 &  $-$4.78 & SG   \\
\Siii& 5669.563 & 14.200 &    0.28  &  $-$5.53 & BBC  \\
\Siii& 5688.817 & 14.186 &    0.13  &  $-$5.50 & BBC  \\
\Siii& 5868.444 & 14.528 &    0.400 &  $-$5.36 & MERL \\
\Siii& 5957.559 & 10.067 & $-$0.301 &  $-$4.84 & SG   \\
\Sii & 6155.134 &  5.619 & $-$0.754 &  $-$3.16 & K07  \\
\Siii& 6347.109 &  8.121 &    0.297 &  $-$5.04 & BBCB \\
\hline
\end{tabular}
\end{center}
The columns give the ion identification,
central wavelength, the excitation potential (in eV) of the lower level, oscillator strength ($\loggf$), 
the Stark damping constant (``appx'' marks lines for which the classical approximation was used), and the reference 
for oscillator strength.\\
AZ -- \citet{AZ}; BBC -- \citet{BBC95}; BBCB -- \citet{BBCB}; LZ €- \citet{LZ71}; KP -€ \citet{KP75}; K07 -- 
http://kurucz.harvard.edu/atoms/1400/; MERL -- \citet{Math01}; SG -- \citet{SG}. Stark damping constants are taken from \vald\
except those marked by asterisk, which are taken from \citet{Wilke03}. 
\end{footnotesize}
\end{table}


\section{Results and discussion}

\subsection{Fundamental atmospheric parameters}

We used an iterative process in which the atmospheric parameters $\teff$, $\logg$, and abundances are gradually improved 
by using different spectroscopic and photometric indicators. 
Using calibration of Str\"omgren indices and Geneva photometry as implemented in the package \templogg \citep{kaiser-2006}, 
as well as simultaneous fit to the observed profiles of $\halpha$ and $\hbeta$ lines, \citet{Joshi} obtained 
the following atmospheric parameters: $\teff=9500$~K, $\logg=3.6$ employing \atl~\citep{a9-1,a9-2} model atmosphere 
computed with scaled solar abundances $\mh=+0.5$~dex. We used this model as starting point in our analysis.
Implementing the iterative procedure described in Sect.~\ref{sec:iter} two sets of parameters were
obtained after five iterations as those providing, on average, a good agreement between observed and predicted spectroscopic 
and photometric observables: $\teff=9300$~K, $\logg=3.5$ and $\teff=9500$~K, $\logg=3.6$.
As will be discussed later, in some cases hotter models provided better fit to certain photometric parameters,
but the ionization balance and hydrogen lines were then subject of noticeable disagreement.

\begin{table}
\caption[ ]{LTE homogeneous atmospheric abundances of \hd.}
\label{habun}
\begin{tabular}{llllll}
\noalign{\smallskip}
\hline
Ion &\multicolumn{3}{c}{This work} & Joshi \etal &  Sun \\
    & t9300g3.5 & t9500g3.6&$n$& t9500g3.6 &\\
\hline
\C   & $-$4.20$\pm$0.15 & $-$4.11$\pm$0.15   &  {\tiny 2}   & $-$4.11$\pm$0.23  & $-$3.65 \\			
\O   & $-$3.81$\pm$0.22 & $-$3.79$\pm$0.22   &  {\tiny 2}   & $-$3.92:          & $-$3.38 \\			
\Na  & $-$4.93$\pm$0.16 & $-$4.82$\pm$0.20   &  {\tiny 2}   & $-$5.09:          & $-$5.87 \\			
\Mgi & $-$3.57$\pm$0.14 & $-$3.45$\pm$0.14   &  {\tiny 5}   & $-$3.58$\pm$0.13  & $-$4.51 \\			
\Mgii& $-$4.29$\pm$0.35 & $-$4.29$\pm$0.35   &  {\tiny 5}   & $-$4.43$\pm$0.27  & $-$4.51 \\			
\Al  & $-$6.21:         & $-$6.25:           &  {\tiny 1}   & $-$6.06:          & $-$5.67 \\		
\Sii & $-$3.73$\pm$0.32 & $-$3.63$\pm$0.32   &  {\tiny 4}   & $-$3.65$\pm$0.33  & $-$4.53 \\		
\Siii& $-$3.94$\pm$0.36 & $-$3.99$\pm$0.34   &  {\tiny 7}   & $-$3.64$\pm$0.47  & $-$4.53 \\			
\Ca  & $-$5.87$\pm$0.09 & $-$5.79$\pm$0.09   &  {\tiny 2}   & $-$5.91           & $-$5.73 \\			
\Ti  & $-$6.46$\pm$0.17 & $-$6.35$\pm$0.17   &  {\tiny 25}  & $-$6.45$\pm$0.16  & $-$7.14 \\			
\V   & $-$8.17$\pm$0.18 & $-$8.07$\pm$0.18   &  {\tiny 2}   & $-$8.14$\pm$0.18  & $-$8.04 \\		
\Cri & $-$3.31$\pm$0.18 & $-$3.19$\pm$0.19   &  {\tiny 101} & $-$3.25$\pm$0.23  & $-$6.40 \\			
\Crii& $-$3.36$\pm$0.22 & $-$3.32$\pm$0.22   &  {\tiny 230} & $-$3.31$\pm$0.24  & $-$6.40 \\			
\Mni & $-$5.93$\pm$0.10 & $-$5.79$\pm$0.10   &  {\tiny 2}   & $-$5.94$\pm$0.11  & $-$6.65 \\			
\Mnii& $-$5.76$\pm$0.24 & $-$5.71$\pm$0.24   &  {\tiny 3}   & $-$5.72$\pm$0.23  & $-$6.65 \\			
\Fei & $-$3.04$\pm$0.16 & $-$2.92$\pm$0.16   &  {\tiny 105} & $-$2.98$\pm$0.20  & $-$4.59 \\			
\Feii& $-$3.12$\pm$0.18 & $-$3.11$\pm$0.18   &  {\tiny 169} & $-$3.01$\pm$0.18  & $-$4.59 \\			
\Co  & $-$5.62:         & $-$5.57:           &  {\tiny 1}   & $-$5.58:          & $-$7.12 \\			
\Ni  & $-$5.33:         & $-$5.21:           &  {\tiny 1}   & $-$5.34:          & $-$5.81 \\		
\Ba  & $-$8.72$\pm$0.25 & $-$8.53$\pm$0.25   &  {\tiny 4}   & $-$8.64:          & $-$9.87 \\			
\Ce  & $-$9.16:         & $-$8.96:           &  {\tiny 1}   & $-$9.10:          & $-$10.46 \\		
\Pr  & $-$9.02$\pm$0.22 & $-$8.91$\pm$0.22   &  {\tiny 5}   & $-$8.87$\pm$0.33  & $-$11.33 \\			
\Nd  & $-$8.51$\pm$0.16 & $-$8.40$\pm$0.17   &  {\tiny 12}  & $-$8.41$\pm$0.16  & $-$10.59 \\			
\Sm  & $-$8.76:         & $-$8.58:           &  {\tiny 1}   & $-$8.70:          & $-$11.03 \\		
\Eu  & $-$8.94$\pm$0.14 & $-$8.72$\pm$0.15   &  {\tiny 2}   & $-$8.85$\pm$0.12  & $-$11.53 \\
\Gd  & $-$8.63$\pm$0.19 & $-$8.47$\pm$0.21   &  {\tiny 2}   & $-$8.72:          & $-$10.92 \\						
\hline
\end{tabular}
Error estimates are based on the internal scattering from the number of analysed lines, $n$. 
Abundance are given in $\log(N_/N_{\rm total})$.
The last column gives the abundances of the solar atmosphere \citep{asplund}.
\end{table}

\begin{figure*}
\epsfig{file=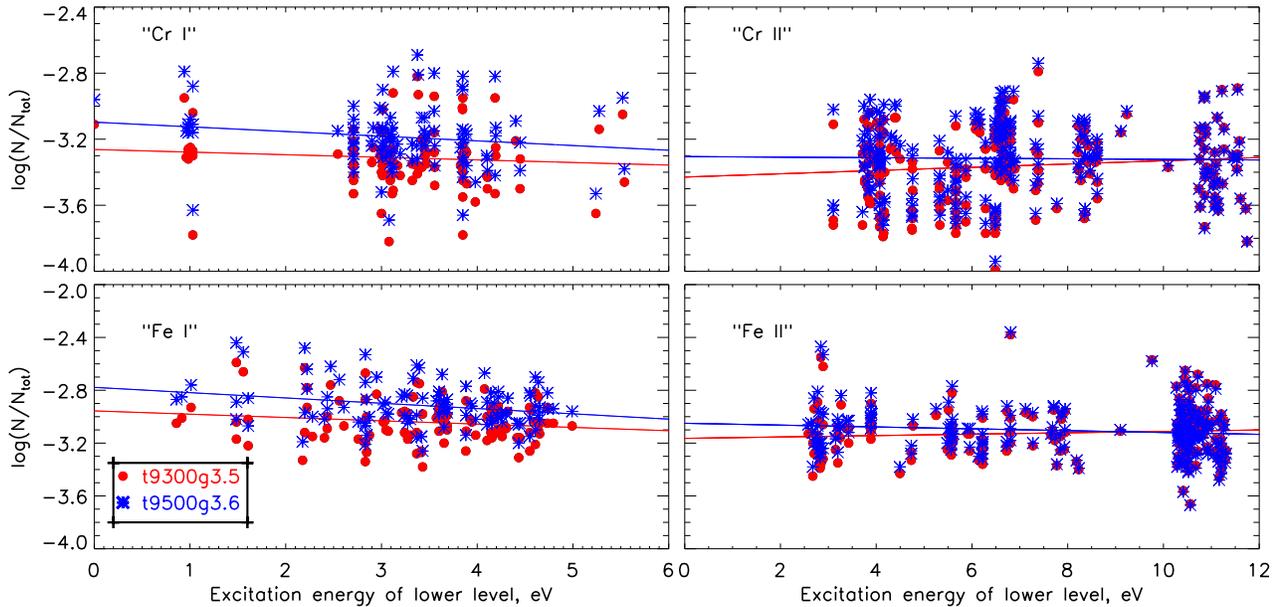,width=1\textwidth}
\caption{Plots of individual abundances for \Cri/\Crii (top) and \Fei/\Feii (bottom) lines 
as a function of excitation energy of the lower level for \hd\ wih two different models: 
red filled circles~--~$\teff=9300$~K, $\logg=3.5$; blue stars~--~$\teff=9500$~K, $\logg=3.6$.}
\label{extbal}
\end{figure*}

The results of abundance analysis shows that the \hd\ is Cr-Fe rich star, and are consistent with those
obtained before by \citet{Joshi}.
Table~\ref{habun} summarizes abundances obtained using two model atmospheres.
The differences between abundances from  \citet{Joshi} and ours for the $\teff=9500$~K, $\logg=3.6$ model
is likely due to the changes in atmospheric structure between solar-scaled abundance model and model atmosphere
computed with realistic chemistry of the star. In addition, the number of lines used for abundance analysis is different
between the two studies which may also influence final results. Nevertheless, the agreement is very good within 
the estimated error bars. As expected, the model with cooler temperature statistically results on lower abundances
as computed atomic lines become stronger with the decrease of $\teff$ in this  particular temperature regime.

Figure~\ref{extbal} illustrates the ionization balance in the atmosphere of \hd\ as derived using lines of
\Fei/\Feii\ and \Cri/\Crii. There is not much difference between the two models and the slopes are
nearly the same, though model with $\teff=9500$~K provides slightly better fit for first ions, while
neutrals seem to prefer $\teff=9300$~K.

\subsection{Stratification of Si and Mg}

Although \hd\ is Cr-Fe rich star, these two elements do not demonstrate any
signature of possible stratification. Abundances obtained from strong and weak lines of Fe and Cr
were found to be nearly the same without noticeable patterns. Since lines with different excitation energies
statistically probe different atmospheric layers, the absence of peculiar abundance patterns among strong and weak
lines clearly shows the absence of strong abundance gradients in the stellar atmosphere, observed in cooler CP stars 
\citep[see, for example, Fig.~3 in ][]{2008CoSka..38..257R}. Nevertheless, we checked possible stratification with the 
{\sc DDAFit} package. We selected 17 lines of the neutral and singly ionised element for Cr and the same amount for Fe.
The range of excitation energies were (0--3.45 eV) for \Cri\ lines, (3.71--11.10 eV) for \Crii\ lines, (0.86--3.39 eV) for
\Fei\ lines, and (2.90--10.68 eV) for \Feii\ lines. The observed intensity varies from 11 to 300 m\AA. Formally, for both 
elements we get a solution with small steep abundance gradient at $\log\tau_{\rm 5000}\approx-1$ which does not exceeds 0.5~dex, 
however, a difference between standard deviation for homogeneous and stratified abundance distributions is insignificant. For both
elements, Cr and Fe, we get a small increase of abundance towards the upper atmosphere 
(except Cr for $\teff=9300$~K, $\logg=3.5$ ). 
Similar distributions are observed in other CP stars of the same temperature range. Cr abundance gradient is very small in 
HD~133792 with $\teff=9400$~K \citep{2006A&A...460..831K} and practically absent in HD~204411 with $\teff=8700$~K \citep{tr2}. 
Recent diffusion calculations \citep{LMHH09,AS10} show that rather strong Cr and Fe abundance gradients with accumulation in 
deeper atmospheric layers should be observed in atmosphere of
a star with $\teff=9300-9500$~K having small magnetic field, and these gradients disappear at much higher temperatures. Strong 
horizontal magnetic field influence the diffusion process resulting in upward abundance gradient after abundance minimum at
$\log\tau_{\rm 5000}\approx-1$. In this case step-function approximation is not valid anymore, 
and its employment leads to smearing the possible gradients. However, no indications exist in favour of the strong horizontal
(toroidal?) component of magnetic field in \hd\ as well as in HD~133792 and HD~204411.    

A detailed inspection of the spectra of \hd\ showed that only two elements are suitable for stratification
analysis: Mg and Si. Both are represented by a sufficient number of strong and weak lines which
are not blended much by nearby lines of others elements. These lines and their current parameters extracted 
from \vald\ are presented in Table~\ref{Tstrat-list}.

\begin{figure}
\epsfig{file=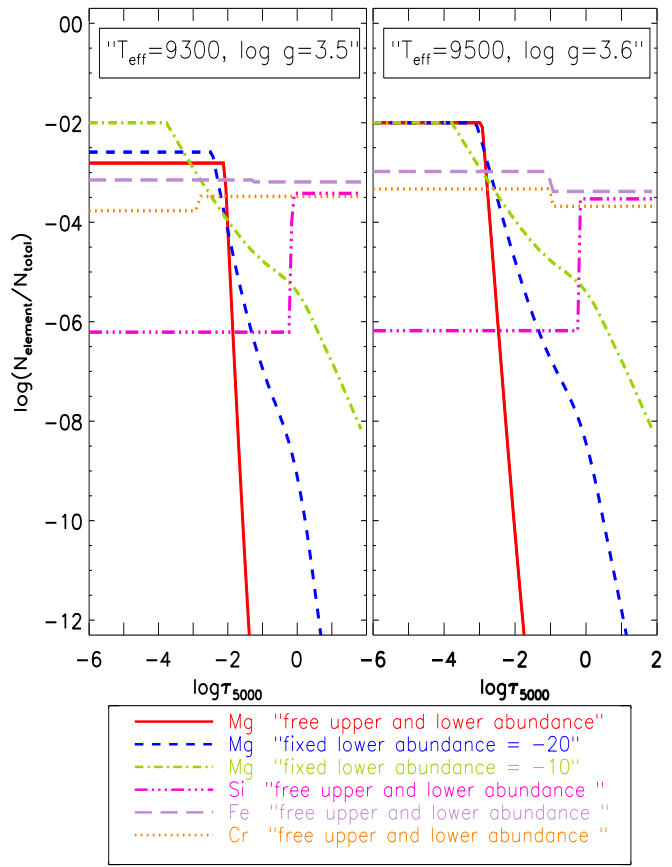,width=.49\textwidth, height=.49\textheight}
\caption{Stratification of Mg and Si in atmosphere of \hd\ derived using two models with $\teff=9300$~K (left panel) 
and $\teff=9500$~K (right panel) and different optimization settings for the step-like stratification profile. 
Full red line~--~optimization of all four parameters of the stratification profile of Mg; 
dashed blue line~--~lower abundance is fixed for Mg at $\log(N/N_{\rm total})=-20$~dex;
dash-dotted green line~--~lower abundance is fixed for Mg at $\log(N/N_{\rm total})=-10$~dex;
dash-three dots magenta line~--~optimization of all four parameters of the stratification profile of Si; 
 purple long-dashes ~--~ optimization of all four parameters of the stratification profile of Fe and 
finally dotted orange line ~--~ optimization of all four parameters of the stratification profile of Cr.}
\label{tauabun}
\end{figure}

\begin{figure}
\epsfig{file=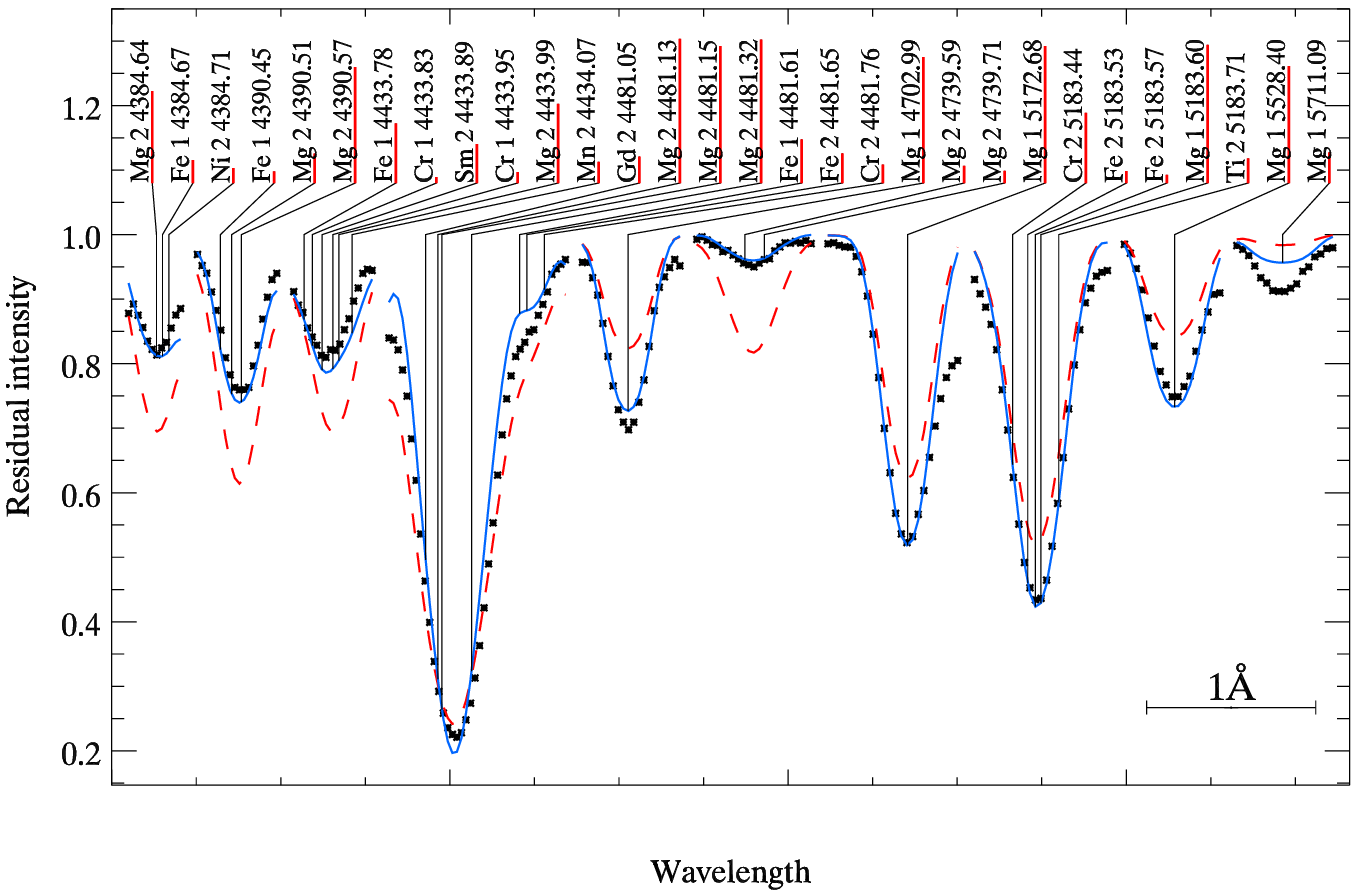,width=.48\textwidth}
\epsfig{file=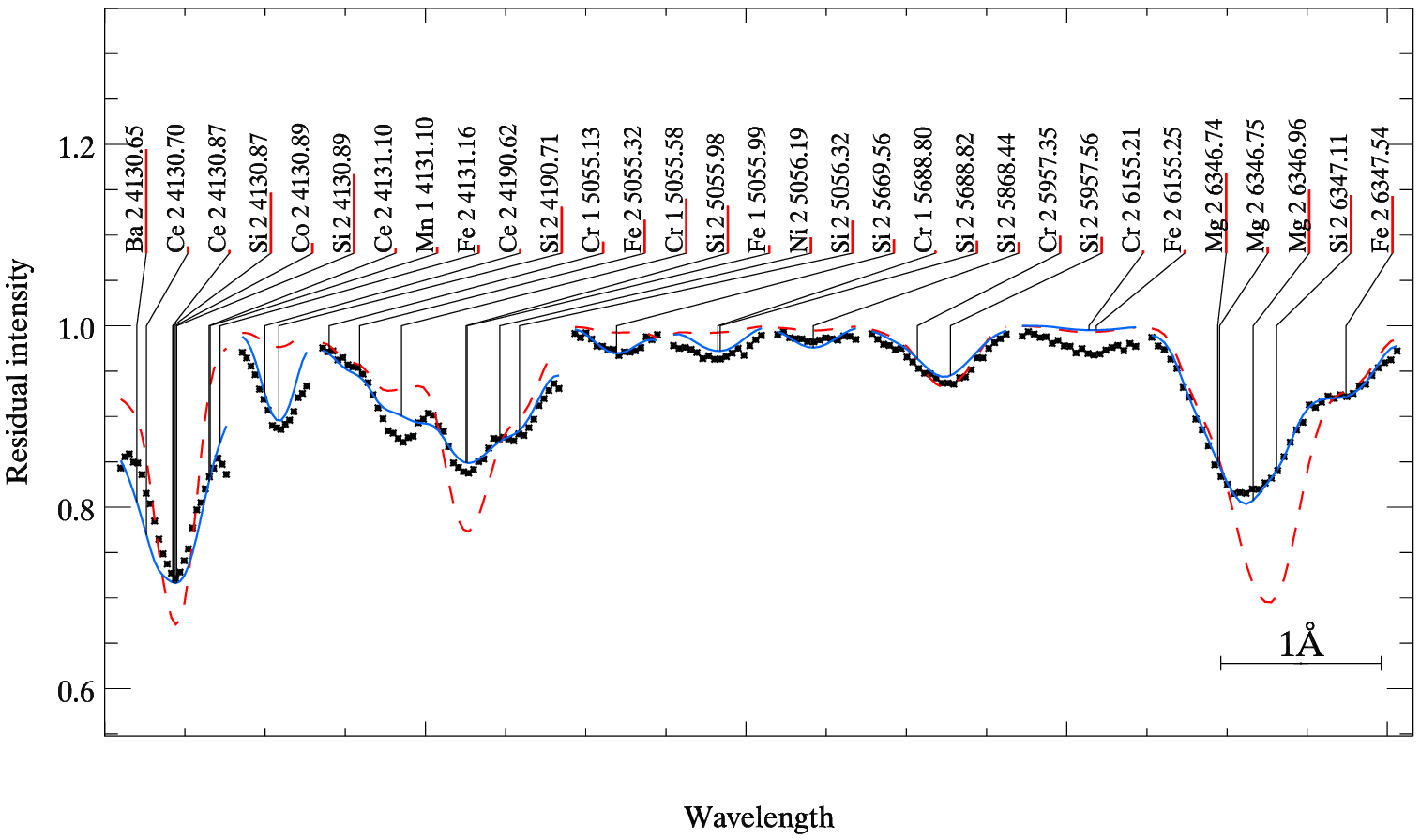,width=.48\textwidth}
\caption{A comparison between the observed and predicted profiles of 
Mg (top panel) and  Si (bottom panel). 
Dotted black line~--~observations; dashed red line~--~synthetic spectra computed with homogeneous
abundances ($\log(N_{\rm Mg}/N_{\rm total})=-3.87$, $\log(N_{Si}/N_{\rm total})=-4.95$) (in case of Si for initial guess we used the homogeneous abundance which gives the best fit to line profiles used for stratification analysis); 
solid blue line~--~synthetic spectra computed with the final stratified abundance distributions
shown by solid red lines in Fig.~\ref{tauabun}.}
\label{Mgstrat}
\end{figure}

Using the \ddafit\ package we derived stratification profiles of Mg and Si for two selected model atmospheres.
Figure~\ref{tauabun} demonstrates run of abundances as a function of atmospheric depth and Fig.~\ref{Mgstrat}
illustrates the resulting best fit to individual line profiles.
We find magnesium to be strongly overabundant in upper atmosphere of \hd\ and under-abundant in
deep layers. The abundance jump is very smooth, occupying a considerable region of atmospheric depths.
It is to note that in case when all four parameters of the assumed step-like abundance distribution were minimized
simultaneously (see Sect.~\ref{sec:strat} for explanations), the solution was always with big error bars for the value 
of the abundance
in deep layers. We thus also tried to derive stratification profile when abundances in these layers were
fixed at constant values, from $\log(N/N_{\rm total})=-20$~dex to $\log(N/N_{\rm total})=-10$~dex
(some examples are presented on Fig.~\ref{tauabun}). However, in all cases the fit to Mg lines did not change much
as well as the common feature of strong abundance gradient between surface and photospheric layers stay still.
It is important to stress here that any values of abundances below $\log\tau_{\rm 5000}\approx 0$ have no strict physical
meaning because these atmospheric layers are not visible to us and thus cannot be modelled by comparing observed and
predicted line profiles. In other words, changing abundances in these layers has no effect on computed line profiles
and hence the true abundances can not be recovered in principle.

\begin{table}
\caption[ ]{Non-LTE abundance corrections associated with each \Mgi\ and \Mgii\ transition used for stratification analysis.  Column 3 and 6 represents NLTE corrections for the star Vega \citep{przybilla}.}
\label{tab1}
\begin{center}
\begin{tabular}{cc|ccc}
\hline
       & $\Delta\log\epsilon$ &        &   $\Delta\log\epsilon $ \\ 
\Mgi\      &        ({\tiny HD103498}) ({\tiny Vega}) &\Mgii\ &  ({\tiny HD103498})  ({\tiny Vega}) \\ \hline
 4702.991  &    $+$0.01 ~~  $+$ 0.02 &  4384.637   &  $-$ 0.00   ~~ .......... \\ 
 5172.684  &    $-$0.03 ~~  $-$ 0.06 &  4390.572   &  $-$ 0.00   ~~ $+$ 0.00\\
 5183.604  &    $-$0.04 ~~  $-$ 0.13 &  4433.988   &  $-$ 0.00   ~~ $+$ 0.00\\
 5528.405  &    $-$0.00 ~~  $-$ 0.03 &  4481.126   &  $-$ 0.05   ~~ $-$ 0.21\\
 5711.088  &    $+$0.01 ~~  ..........& 4739.593   &  $+$ 0.00  ~~ .......... \\
\hline
\end{tabular}
\end{center}
\end{table}

In addition we find that none of the Mg lines used for stratification analysis is a subject of strong deviations from the LTE
which could affect final results.
Since \detail/\surface\ package can only deal with homogeneous abundances, 
we used the original model atmosphere from \citet{Joshi} in order to estimate respective non-LTE abundance corrections
$\Delta \log \epsilon = \log \epsilon _{NLTE} - \log \epsilon_{LTE}$, which are presented in Table~\ref{tab1}.
The largest non-LTE abundance correction was found for the \Mgii~$4481.126$~\AA\ line 
and is  $-0.05$~dex.

Stratification profile of Si was found to be similar to those frequently found in cooler CP stars, i.e. with
accumulation in photospheric layers. Its shape
did not change much between $\teff=9300$~K and $\teff=9500$~K models, only the abundance jump became more steep
for the cooler one.

The fact that Mg stratification is opposite to those of Si, i.e. that the Mg is pushed up by radiation 
much more effectively is, of course, a very important result for the modern models of particle diffusion
in stellar atmospheres, and the same behaviour of Mg was already found in another peculiar star
HD~204411 \citep[see Fig.~7 in][]{tr2}.

The obtained Mg distribution again can be explained qualitively by diffusion in the presence of very strong horizontal 
magnetic field \citep[see Fig.~3 in][]{AS07}. In principle, Si gradients are also predicted by diffusion models,
however, neither \citet{AS07,AS10} nor \citet{LMHH09} predict strong Si overabundance at the photospheric layers, that is commonly 
observed in all stars investigated up to now. In all diffusion calculation maximum of Si abundance is solar or slightly below
the solar value.  

Formal distributions of Fe and Cr are also shown on Fig.~\ref{tauabun}, 
but because of relatively small abundance jumps ($\leqslant0.5$~dex) their distributions 
can be safely described as uniform (within the errors of standard deviations between observed and
predicted line profiles). Therefore, even if present, such weak abundance gradients of Fe and Cr would
in no way affect photometric and spectroscopic observables compared to the models computed with simple homogeneous
distributions assumed in the present work.

\subsection{Energy distribution and hydrogen line profiles}

\begin{figure}
\epsfig{file=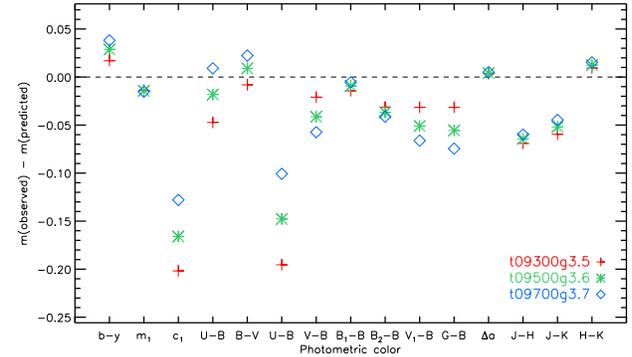,width=.48\textwidth}
\caption{Difference between observed and predicted colors of different photometric systems.
Theoretical colors were reddened applying $E(B-V)=0.06$.}
\label{fig:colors}
\end{figure}

Unfortunately, the lack of observed spectrophotometric or low resolution spectroscopic
energy distributions for \hd\ did not allow us to uniquely derive its effective temperature.
We therefore only attempted to use available photometric observations in different systems.
Figure~\ref{fig:colors} illustrates a difference between observed and computed photometric parameters
for three atmospheric models. 
All photometric parameters were calculated using modified computer codes by 
\citet{a9-2}, which take into account transmission curves of 
individual photometric filters, mirror reflectivity and a photomultiplier 
response function. From the HIPPARCOS parallax of $\pi=3.37\pm0.56$~mas \citep{leeuwen}
the distance to the star is $d=297$~pc, which results in redenning value of $E(B-V)=0.06$ applying
extinction maps published by \citet{2005AJ....130..659A}.
In general, the two final models with $\teff=9300$~K and $\teff=9500$~K
are in reasonable agreement with observations except Str\"omgren $c_{\rm 1}$ and Geneva $U-B$ for which
we find the largest deviation of about $0.2$~mag. As $c_{\rm 1}$ measures the amplitude of the Balmer jump,
one would suppose that the true temperature or gravity of the star should be much higher than adopted in the present study.
Indeed, increasing $\teff$ up to $9700$~K improves the fit to the $c_{\rm 1}$ significantly. However, in this
case the fit to hydrogen lines become much worse. Last but not least, the temperature sensitive index $b-y$
clearly requires lower temperatures. The same concerns gravity of the star: increasing $\logg$
also helps to get close to the observed value of $c_{\rm 1}$, but then we find 
disagreement between observed and predicted hydrogen lines as well as in ionization balance 
between \Fei/\Feii\ and \Cri/\Crii.

\begin{figure*}
\epsfig{file=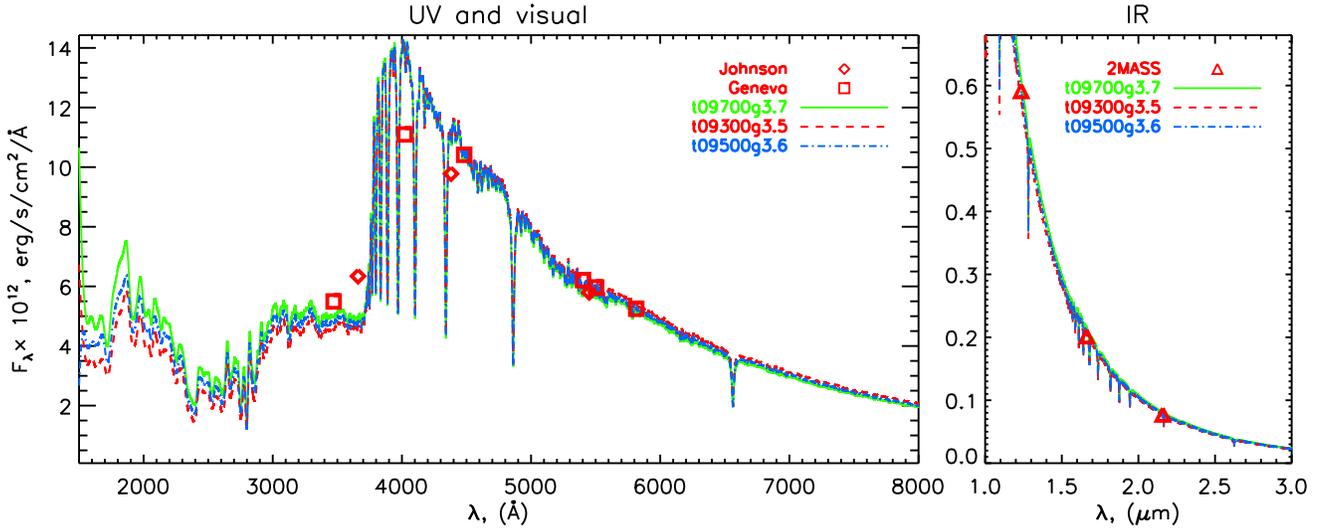,width=1\textwidth}
\caption{Comparison between observed photometric parameters calibrated in absolute units and theoretical
fluxes convolved with FWHM=$5$~\AA\ Gaussian. Theoretical fluxes were reddened applying $E(B-V)=0.06$.}
\label{fig:flux}
\end{figure*}

To better understand the problem, we plotted observed photometric parameters transformed
to absolute fluxes and compared them with model predictions on Fig.~\ref{fig:flux}.
To obtained fluxes in physical units, we used calibrations by  \citet{rufenernicolet} for the Geneva,
\citet{vander} for the 2MASS, and \citet{bessel} for the Johnson photometry. As expected, infrared
fluxes are not sensitive very much to the temperature and gravity of models, but the strong
deviation in Johnson and Geneva ultraviolet $U$ indices is clearly visible.

\begin{figure*}
\epsfig{file=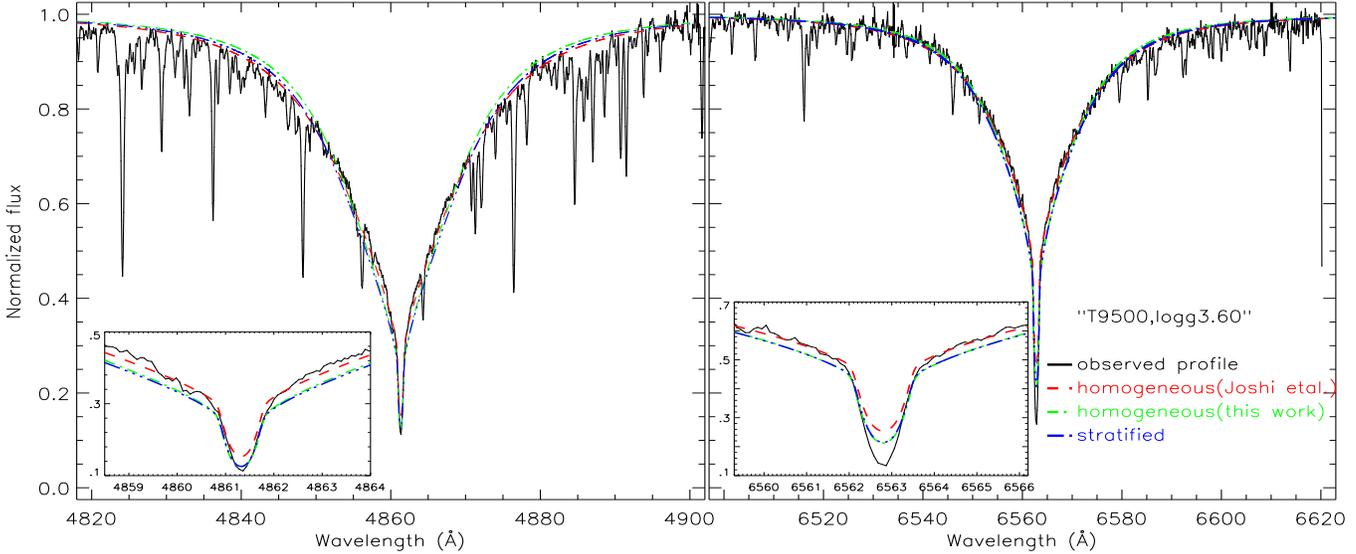,width=1.05\textwidth,  height=.45\hsize}
\caption{Comparison between the observed $\hbeta$ (left panel),  $\halpha$ (right panel)  observed and synthetic line profiles. Solid black line ~--~observed profile; dashed red line~--~synthetic spectra cumputed with homogeneous abundances \citep{Joshi}; dash-dotted green line~--~synthetic spectra cumputed with homogeneous abundances (this work); dash-three dots blue line~--~synthetic spectra computed with stratified abundances. In both cases the final adopted model corresponds to $\teff=9500$~K, $\logg=3.6$}. The inset shows the zoomed part of the profiles around the core of line.
\label{halpha}
\end{figure*}

The fit to the Balmer $\halpha$ and $\hbeta$ lines is presented on Fig.~\ref{halpha}. Accounting for the stratification
clearly helps to improve the fit to the observed line profiles compared to the spectra computed with only homogeneous
abundances, but both spectra fail to fit the transition region between cores and wings of lines where theoretical
profiles are systematically too much wide. The same holds true for the cooler model with $\teff=9300$~K (not shown
on the figure for the better view). Still, the solar-scaled abundance model used by \citet{Joshi} seems to
provide a best fit among all the models considered in this study, however one should not forget that this model
is inconsistent with the true abundance pattern of the star: \citet{Joshi} used \atl\ model with metallicity
$\mh=+0.5$~dex, but the true iron overabundance of the star is $\approx+1.5$~dex compared to the Sun. Obviously, using now
respective homogeneous model with $\mh=+1.5$~dex would immediately result in wider profiles of hydrogen lines 
(due to increase in number of free electrons and thus Stark broadening) no longer providing a fit to line profiles.

From the analysis presented above its clear that there is a fundamental disagreement between temperatures of the star
suggested by some photometric indicators and hydrogen line profiles. Fitting energy distribution  require systematically hotter 
temperature than those followed from fitting the $\halpha$ and $\hbeta$ lines. We thus speculate that there could be few
reasons for this that should be checked in future investigations. First of all, except Mg and Si other elements can be stratified
in the atmosphere of \hd. If so and (as usually found) the stratification profiles are similar to
that of Si, then the concentrations of free electrons in upper atmospheric layers could becomes smaller 
than the one of a present study, thus resulting in narrower
profiles of hydrogen lines. Unfortunately, as stated above, no good sets of lines for accurate stratification analysis
could be found for elements over than Mg and Si in the spectra of \hd. Also, the normalization of hydrogen lines
in echelle spectra is a complicated task and some systematic uncertainties may still influence fitting results.

Finally, comparison of observed photometry calibrated in absolute units and model fluxes allowed us to estimate
the radius of the star. With parallax $\pi=3.37\pm0.56$~mas we obtained
$R=(4.56\pm0.77)\Rsun$ for $\teff=9300$~K, $\logg=3.5$ model and $R=(4.39\pm0.75)\Rsun$ for $\teff=9500$~K, $\logg=3.6$ model respectively.

\section{Conclusions}
In this work we carried out detailed atmospheric abundance and stratification analysis of the chemically peculiar
A0 star \hd, which is an extension of analysis presented in \citet{Joshi}.
Using available spectroscopic and photometric parameters, as well as up-to-date model atmosphere 
and spectra synthesis codes we attempted to construct a self-consistent model atmosphere of the star and to
determine fundamental atmospheric parameters iteratively by comparing various observables with the
model predictions. Our main results are summarized in following:
\begin{itemize}
\item
We find that, generally, two model atmospheres computed with individual and stratified
abundances provide reasonable fit simultaneously to spectroscopic and photometric indicators:
$\teff=9300$~K, $\logg=3.5$ and $\teff=9500$~K, $\logg=3.6$. The later model has the same parameters
as the one used in \citet{Joshi}, but, in contrast, includes realistic chemistry of the star.
\item
The abundance analysis demonstrate strong overabundance of Fe and Cr relative to the Sun, and
our results are in excellent agreement with \citet{Joshi}.
\item  
Using simplified step-function approximation of stratification profile we derived depth-dependent distributions
of Mg and Si. We show that Mg has a large abundance gradient in the atmosphere of \hd, with accumulation
of Mg ions in the uppermost atmospheric layers. Distribution of Si demonstrates opposite behaviour with
accumulation of Si ions in deep atmospheric layers. This indicates that Mg is pushed up by radiative forces
much more efficiently than Si. These empirical findings are of certain interest for the modern theory
of particle diffusion in stellar atmospheres.
\item
 Although the star is Cr-Fe rich, these two elements do not show any signature of noticeable stratification
in its atmosphere.
\item
We find that none of Mg transitions used for stratification analysis is a subject of noticeable non-LTE effects.
The largest non-LTE abundance correction was found for the \Mgii $4481.126$~\AA\ line 
and is  $-0.05$~dex.
\item
Comparing observed photometry in Str\"omgren, Geneva, 2MASS, and Johnson filters transformed to physical
units with model fluxes we estimated the radius of \hd\ to be
$R=4.56\pm0.77~\Rsun$ for $\teff=9300$~K, $\logg=3.5$ model and $R=4.39\pm0.75~\Rsun$ for $\teff=9500$~K, $\logg=3.6$ model respectively.
\item
Comparison of observed photometric parameters and hydrogen line profiles for different model atmospheres
showed that such indices as $c_{\rm 1}$ and $U-B$ systematically require hotter models and $\halpha$, $\hbeta$, and
$b-y$ index cooler models relative to the initially assumed temperature $\teff=9500$~K. Unfortunately, the lack
of suitable spectrophotometric or low resolution spectroscopic observations in absolute units forbids
us to uniquely determine the $\teff$ of the star.
\end{itemize}

\section*{Acknowledgments}
This work was supported by  Deutsche Forschungsgemeinschaft (DFG) Research Grant RE1664/7-1 to DS. TR acknowledges partial support from Presidium RAS programme, and
from the Russian Federal Agency on Science and Innovation (No. 02.740.11.0247). OK is a Royal Swedish Academy of Sciences Research Fellow supported by grants from the Knut and Alice Wallenberg Foundation and the Swedish Research Council. CPP indebted to Dr. K. Butler and Dr. T. Morel for  kindly providing the codes (\detail/\surface) for NLTE analyses as well as for valuable discussions. C.P.P also wish to sincerely thank Dr. B. B. Sanwal, Dr. U. S. Chaubey, Dr. S. Bisht and Prof. R. Sagar for their kind motivation and encouragement. We also wish to thank the anonymous referee for useful comments that helped improving the presentation of this paper.\\
We acknowledge the use of Vienna Atomic Line Database (VALD), the SIMBAD astronomical database and the NASA's ADS. 
This publication makes use of data products from the Two Micron All Sky Survey, which is a joint project of the University of Massachusetts and the Infrared Processing and Analysis Center/California Institute of Technology, funded by the National Aeronautics and Space Administration and the National Science Foundation.

\label{lastpage}
\end{document}